\documentstyle[preprint,eqsecnum,aps]{revtex}
\tighten
\begin{document}
\draft

\title{STRING DYNAMICS IN COSMOLOGICAL
AND BLACK HOLE BACKGROUNDS:
THE NULL STRING EXPANSION}

\author{C. O. LOUSTO}
\address{Department of Physics, University of Utah\\
Salt Lake City, UT 84112, USA}
\author{N. S\'ANCHEZ}
\address{Observatoire de Paris, DEMIRM,
UA 336 Laboratoire associ\' e au CNRS,
\\ 61, Avenue de l'Observatoire, F-75014 Paris, France}
\date{\today}
\maketitle

\begin{abstract}

We study the classical dynamics of a bosonic string
in the $D$--dimensional flat Friedmann--Robertson--Walker
and Schwarzschild backgrounds. We make a perturbative development
in the string coordinates around a {\it null} string configuration;
the background geometry is taken into account exactly. In the
cosmological case we uncouple and solve the first order fluctuations;
the string time evolution with the conformal gauge world-sheet
$\tau$--coordinate is given by
$X^0(\sigma, \tau)=q(\sigma)\tau^{1\over1+2\beta}+c^2B^0(\sigma,\tau)+
\cdots$, $B^0(\sigma,\tau)=\sum_k b_k(\sigma)\tau^k$ where $b_k(\sigma)$
are given by Eqs.\ (\ref{bks}),
and $\beta$ is the exponent of the conformal factor in the
Friedmann--Robertson--Walker metric, i.e. $R\sim\eta^\beta$.
The string proper size, at first order in the fluctuations,
grows like the conformal factor $R(\eta)$ and the
string energy--momentum tensor corresponds to that of
a null fluid. For a string in the black hole background, we
study the planar case, but keep the dimensionality
of the spacetime $D$ generic. In the null string expansion, the
radial, azimuthal, and time coordinates $(r,\phi,t)$ are
$r=\sum_n A^1_{n}(\sigma)(-\tau)^{2n/(D+1)}~,$
$\phi=\sum_n A^3_{n}(\sigma)(-\tau)^{(D-5+2n)/(D+1)}~,$ and
$t=\sum_n A^0_{n}(\sigma)(-\tau)^{1+2n(D-3)/(D+1)}~.$
The first terms of the series represent a
{\it generic} approach to the Schwarzschild
singularity at $r=0$. First and higher
order string perturbations contribute with
higher powers of $\tau$. The integrated string
energy-momentum tensor corresponds
to that of a null fluid in $D-1$ dimensions.
As the string approaches the $r=0$ singularity its proper size grows
indefinitely like
$\sim(-\tau)^{-(D-3)/(D+1)}$. We end the paper giving
three particular exact
string solutions inside the black hole. They represent respectively
straight strings across the origin, twisted, and rigidly rotating
strings.

\end{abstract}

\section{Introduction}

The investigation of strings in curved spacetime is currently the
best framework to study the physics of gravitation in the context
of string theory. The study of string propagation in curved spacetimes
provides new insights with respect to string theory in flat, Minkowski
spacetime (and with respect to quantum fields in curved spacetime)
[See for example Refs.\ \cite{GSW87,BD82,dVS93,SZ96}.]
The results of this study apply to fundamental strings, as well as
to cosmic strings, which behave essentially in a classical way.

The string equations of motion and constraints in curved spacetime
are highly non linear and, in general, non exactly solvable. Different
methods are available to solve this system: The string
perturbation approach\cite{VS87},
the $\tau$--expansion method\cite{GSV91}
(which provides exact local solutions for any background), the null
string approach\cite{VN92}, and the construction of global
solutions (by solitonic and inverse scattering methods, for instance),
which allowed to uncover the new feature of multistring solutions
\cite{CVMS94,VLS94,VE94,LS94}.
The expansion methods are described and classified by using the world
sheet velocity of light, $c$, as an expansion parameter, in Sec.\ II
below.

An approximate but general method is the expansion around the center
of mass solution of the string (satisfying the geodesic
equation of motion), the world sheet time $\tau$ being identified
with the proper time associated to the center of mass trajectory.
In this paper we are concerned with another general approximation
method: The null string approach. In this approach, the string
equations of motion and constraints are systematically expanded in
powers of the parameter $c$ (the speed of light in the world sheet).
This corresponds to a small string tension expansion. To zeroth
order the string is effectively equivalent to a continuous 
line--beam of massless particles labeled by the world sheet spatial
parameter $\sigma$. The points on the ``null'' string do not interact
among them, but they interact with the gravitational background.
Note the extended character of the zeroth order solution in the
null string approach, as opposed to the perturbation approach in
which the zeroth order is given by the string's center of mass
point--like approximation

We also study fluctuations around the null string configuration,
which naturally appears as an expansion in powers of $\tau$ with
precisely defined $\sigma$--dependent coefficients. The string
coordinates are expressed as
\begin{equation}
X^A(\sigma, \tau)=A^A(\sigma, \tau)+c^2B^A(\sigma, \tau)+
c^4C^A(\sigma, \tau)+...\label{2.5'}
\end{equation}
the zeroth order $A^A(\sigma, \tau)$, first and second order
fluctuations $B^A(\sigma, \tau),\,C^A(\sigma, \tau)$, satisfy
Eqs.\ (\ref{2.3})--(\ref{2.4}) as described in Sec.\ II.
This null string approach is a powerful description for strings
in the strong gravitational field regime, as is the case of
black hole backgrounds near the $r=0$ singularity and in
inflationary cosmological spacetimes, at late times, where
string instabilities develop. The string perturbation method
around the string center of mass allows to detect the
emergence of instabilities\cite{SV90,LS93,LS94}, but is unable to
describe the highly unstable regime\cite{GSV91}.
We apply this null string expansion to
cosmological spacetimes and to strings in the interior of black
holes, falling into the singularity at $r=0$.
In Friedmann--Robertson--Walker cosmological
spacetimes with conformal expansion factor $R\propto\eta^\beta$,
$\eta$ being the conformal gauge world sheet time,
the null string evolution can be expressed as
\begin{equation}
X^0(\sigma, \tau)=q(\sigma)\tau^{1\over1+2\beta}+c^2
B^0(\sigma,\tau)+...
\end{equation}
where
\begin{equation}
B^0(\sigma,\tau)=\sum_k b_k(\sigma)\tau^k~,
\quad
k=2,2\pm\alpha/\beta,k_{\pm}~.\nonumber
\end{equation}
The spatial coordinates $X^i(\sigma, \tau)$ are given by
Eqs.\ (\ref{c.5}) and (\ref{eqbi}) in Sec.\ III.

The proper size of the string grows proportional to the expansion
factor $R(\eta)$ and the string energy momentum tensor corresponds
to that of a null fluid.

In black hole spacetimes we find the null string expansion inside
the Schwarzschild black hole. I is given by
\begin{equation}
t\dot=X^0(\sigma,\tau)=\sum_{n=1}^\infty A^0_{n}(\sigma)
(-\tau)^{1+2n(D-3)/(D+1)}+c^2B^0(\sigma,\tau)+\cdots
\end{equation}
\begin{equation}
r\dot=X^1(\sigma,\tau)=\sum_{n=1}^\infty A^1_{n}(\sigma)
(-\tau)^{2n/(D+1)}+c^2B^1(\sigma,\tau)+\cdots
\end{equation}
\begin{equation}
\phi\dot=X^3(\sigma,\tau)=\sum_{n=1}^\infty A^3_{n}(\sigma)
(-\tau)^{(D-5+2n)/(D+1)}+c^2B^3(\sigma,\tau)+\cdots
\end{equation}
and we have taken, for simplicity, the string in the equatorial plane,
i. e.
\begin{equation}
\theta^i\dot=A^i=\pi/2 {\text{ for}}~~ i=2,4,5,..., D-1.
\end{equation}

The $\tau$--dependence exhibited above represents a generic behavior
near the $r\to0$ singularity $(\tau\to0)$, and generalizes to
$D$--dimensions the case recently analyzed in Ref.\ \cite{VE95}.
Higher order fluctuations $c^2B^\mu$, $c^4C^\mu$, and so on,
contribute with  increasing orders of $\tau$.

The proper string size grows indefinitely as the string approaches
the $r\to0$ singularity as
\begin{equation}
\left({dl\over d\sigma}\right)\to\left({R_S\over r}\right)^{(D-3)/2}
{X^0}'(\sigma,\tau=0)\to(-\tau)^{-(D-3)/(D+1)}~.
\end{equation}

The integrated string energy--momentum tensor for $r\to0$ behaves as
\begin{equation}
\Theta^r_r\to-{L\over2\pi\alpha'r}=-\Theta^\phi_\phi~,~~\Theta^t_t\to0~.
\end{equation}
$L$ being the string orbit angular momentum. In the infalling towards
the $r\to0$ singularity, the string behaves as a null fluid in
$(D-1)$--dimensions; in particular, a planar string behaves as a two
(spatially) dimensional null fluid. The emergence of string instabilities
in Schwarzschild and Reissner--Nordstr\"om spacetimes, and the growing
of the string proper size near the black hole singularity was first found
in Ref.\ \cite{LS93} using the perturbative expansion method around the
string center of mass (although such expansion does not allow to describe
the highly unstable regime and the full approach to the $r\to0$
singularity).

This paper is organized as follows: In Sec.\ II we classify
the different expansion methods to solve the string dynamics in 
curved spacetimes in terms of the parameter $c$ (or equivalently,
in terms of the ratio between $\dot X^A$ and $c{X^A}'$, the $\tau$
and $\sigma$ derivatives of the string coordinates), with particular
focusing on the null string expansion (See also Table II). In Sec.\ III
we solve the null string expansion in cosmological spacetimes.
In Sec.\ IV we find the null string expansion near the Schwarzschild
$r\to0$ singularity. In Sec.\ V we find particular exact solutions
inside the black hole event horizon describing straight string across
$r=0$, twisted and rigidly rotating strings. Sec.\ VI summarizes our
conclusions.

\section{String dynamics in gravitational backgrounds}

The action of a bosonic string in a $D$--dimensional curved
manifold endowed with
a metric $G_{AB}$ ($0\leq A, B\leq D-1$) is given by \cite{L84,FT85}
\begin{equation}
S=-T_0\int{d\sigma d\tau\sqrt{-\det g_{\mu\nu}}}
\label{2.1}
\end{equation}
where $g_{\mu\nu}=G_{AB}(X)\partial_{\mu}X^A\partial_{\nu}X^B$
is the two dimensional world--sheet metric 
($0\leq\mu, \nu \leq1$) and $T_0=1/(2\pi\alpha')$ is the string tension.

By use of the reparametrization invariance of the world--sheet one can take
the conformal gauge, i.e. $g_{\mu\nu}=\rho(\sigma, \tau)\eta_{\mu\nu}$,
where $\eta_{\mu\nu}$ is the 
two-dimensional Minkowskian metric. In this gauge, the classical equations of 
motion derived from the action (\ref{2.1}) read
\begin{equation}
\partial^2_{\tau}X^A-c^2\partial_{\sigma}X^A + \Gamma^A_{BC}\left[
\partial_{\tau}X^B\partial_{\tau}X^C-c^2\partial_{\sigma}X^B
\partial_{\sigma}X^C\right]=0~,
\label{2.2}
\end{equation}
where $\Gamma^A_{BC}$ are the Christoffel symbols associated to the metric
$G_{AB}$ and we have introduced the velocity of wave propagation along
the string (velocity of light), $c$, as the second fundamental constant 
after the string tension.

Variation of action (\ref{2.1}) with respect to the world--sheet variables
yields the two non--trivial constraints
\begin{equation}
\partial_{\tau}X^A\partial_{\sigma}X^B G_{AB}=0~, \label{2.3}
\end{equation}
\begin{equation}
[\partial_{\tau}X^A\partial_{\tau}X^B+c^2\partial_{\sigma}X^A\partial_{\sigma}
X^B]G_{AB}=0~.     \label{2.4}
\end{equation}

de Vega and S\'anchez \cite{VS87} have proposed a method to solve, both
classically and quantum mechanically, the equations of motion and 
constraints of strings in curved spacetimes, i. e. Eqs. 
(\ref{2.2})-(\ref{2.4}).
This method treats the spacetime geometry exactly and considers the string
excitations small as compared to the energy
scales associated to the background gravitational metric. This method is
particularly well suited to study strings 
in strong gravitational fields such us in black holes and in cosmological
scenarios as
opposed to the usual treatment in flat spacetime. The starting point is to 
consider a particular solution to the equations of motion and constraints,
$X_{\text{part}}^A(\sigma, \tau)=A^A(\sigma, \tau)$ and then successively
studying first, second and
higher order fluctuations around this particular solution; to be denoted
respectively as $B^A, C^A,$ etc.

In Ref.\ \cite{VN92} have been raised the interesting possibility
of using $c$, the world--sheet light velocity, as the expansion parameter
in the above development. This approach allow us to classify the different 
solutions proposed in the literature:

\begin{itemize}
\item i) $c\ll 1$ or equivalently, as can be seen from the field and
constraint equations (\ref{2.2})-(\ref{2.4}),
$\partial_{\tau}A^A\gg c\partial_{\sigma}A^A$
(holding also this inequality for second derivatives).
\par\noindent
\item ii) $c=1$ and thus all derivatives and other terms are, in principle,
relevant in Eqs. (\ref{2.2})-(\ref{2.4}).
\par\noindent
\item iii) $c\gg 1$ or equivalently,
$c\partial_{\sigma}A^A\gg \partial_{\tau} A^A$
(holding also for second derivatives).
\par\noindent
\item iv) There can be, of course, several other possibilities
such as hybrid starting solutions
that for some components of $A^A$ (and eventually $B^A$, $C^A$, etc)
fulfill one of the cases i), ii) or iii) and for other components a
different case. \end{itemize}

More explicitly, we have

\noindent
{\bf Case} i) $c\ll 1$:
The appropriate development reads
\begin{equation}
X^A(\sigma, \tau)=A^A(\sigma, \tau)+c^2B^A(\sigma, \tau)+
c^4C^A(\sigma, \tau)+...\label{2.5}
\end{equation}
Plugging this into Eqs. (\ref{2.2})-(\ref{2.4}) one obtains
\cite{VN92}, to zeroth order in $c$
(since now on we adopt the notation $\dot{~}=\partial_{\tau}$
and $' =\partial_{\sigma}$)
\begin{equation}
\ddot A^A+\Gamma_{BC}^A\dot A^B\dot A^C=0~, \label{2.6}
\end{equation}
\begin{equation}
\dot A^B\dot A^CG_{BC}=0~, \label{2.7}
\end{equation}
\begin{equation}
\dot A^B A'^CG_{BC}=0~. \label{2.8}
\end{equation}

Eq.\ (\ref{2.6}) is the geodesic path followed by each point of 
the string. Eq.\ (\ref{2.7}) means that this is a null geodesic and
Eq.\ (\ref{2.8})
equation states that the velocity is perpendicular to the string.

First order fluctuations around this particular solution $A(\sigma, \tau)$
can be obtained by retaining terms of order $c^2$ in Eqs
(\ref{2.2})-(\ref{2.4})
\begin{equation}
\ddot B^A+2\Gamma_{BC}^A\dot A^B\dot B^C+\Gamma_{BC,D}^A\dot A^B\dot A^C
B^D=A''^A+\Gamma_{BC}^AA'^BA'^C~, \label{2.9}
\end{equation}
\begin{equation}
(2\dot A^B\dot B^C+A'^BA'^C)G_{BC}+G_{BC,D}\dot A^B\dot A^C B^D=0~,
\label{2.10}
\end{equation}
\begin{equation}
(\dot B^BA'^C+\dot A^BB'^C)G_{BC}+G_{BC,D}\dot A^B A'^C B^D=0~.
\label{2.11}
\end{equation}
Higher order corrections can thus be systematically obtained.
The interpretation of these equations is that they give the
high energy (compared to the Planck energy) string behavior,
corresponding to the limit of vanishing tension (and thus small $c$).

\noindent
{\bf Case} ii) $c=1$, and thus all derivatives and other
terms are, in principle,
relevant in Eqs. (\ref{2.2})-(\ref{2.4}). One can thus make developments
around the center--of--mass motion as a physically appealing starting
solution. In fact, de Vega and S\'anchez \cite{VS87} have proposed in
his original approach
\begin{equation}
X^A=A^A(\sigma, \tau)+B^A(\sigma, \tau)+C^A(\sigma, \tau)+...~, \label{2.12}
\end{equation}
where, $A^A(\tau)$ follows the geodesic equation
\begin{equation}
\ddot A^A+\Gamma_{BC}^A\dot A^B\dot A^C=0~,\label{2.13}
\end{equation}
and the constraint
\begin{equation}
\dot A^B\dot A^CG_{BC}=0~,\label{2.14}
\end{equation}
means that these geodesics are null.

First order fluctuation equations now read
\begin{equation}
\ddot B^A-B''^A+2\Gamma_{BC}^A\dot A^B\dot B^C+\Gamma_{BC,D}^A\dot A^B
\dot A^CB^D=0~,\label{2.15}
\end{equation}
and
\begin{equation}
2G_{BC}\dot A^B\dot B^C+G_{BC,D}\dot A^B\dot A^CB^D=0~,\label{2.16}
\end{equation}
\begin{equation}
G_{BC}\dot A^BB'^C=0~.\label{2.17}
\end{equation}
Higher order corrections can be in this way systematically considered.
Still, a Fourier  series decomposition of the first order fluctuations 
can be made \cite{VS87}, i. e. $B^A(\sigma, \tau)=\Sigma_n\exp{(in\sigma)}
\eta_n^A(\tau)$. Also, a decomposition in purely left
$A^A(\sigma+\tau)$ or right movers
$A^A(\sigma-\tau)$ can be used as a starting exact solution.

\noindent
{\bf Case} iii) $c\gg 1$ or equivalently,
$c\partial_{\sigma}A^A\gg \partial_{\tau}
A^A$ (holding also for second derivatives).
The appropriate development reads in this case
\begin{equation}
X^A(\sigma, \tau)=A^A(\sigma, \tau)+{1\over c^2}B^A(\sigma, \tau)+
{1\over c^4}C^A(\sigma, \tau)+...\label{2.18}
\end{equation}   
Plugging this expression into Eqs. (\ref{2.2})-(\ref{2.4})
yields a set of equations
that is equivalent to (\ref{2.6})-(\ref{2.11}) upon the substitutions
$\tau\leftrightarrow\sigma$ and $c\to c^{-1}$.
The interpretation of this approximation
is now exactly the opposite to case i), i. e. here the zeroth order dependence
on $\sigma$ overwhelms that on $\tau$ and thus represent very low energy 
strings, frozen at first approximation. Solutions to this case can be 
obtained from solutions to case i) by making the above mentioned substitutions.
These represent, at zeroth order, stationary solutions as opposed to the
dynamical ones. While solutions to case i) are appropriate
to describe strings in strong gravitational fields, for instance,
near black hole singularities, i.e. $R\to0$,
solutions to the case iii) are rather appropriate for strings far away from
black holes, i.e. $R\to\infty$. We thus have an approximate duality here,
represented by the transformation $\tau\leftrightarrow\sigma$ and
$R\leftrightarrow R^{-1}$, which map the solutions to the cases i) and
iii) into one another.

\noindent
{\bf Case} iv): This corresponds to an hybrid case. The circular
string ansatz is an example of this situation:
\begin{equation}
X^0=X^0(\tau)~,~~~R=R(\tau)~,~~~\psi=\psi(\sigma)~,~~\theta=\pi/2~.
\label{3.13}
\end{equation}
Clearly $X^0$ and $R$ coordinates follow criterion i) while $\psi$
follows case iii) and $\theta$ could be assigned to case ii).
The first order fluctuations around this starting solution have been
studied in Ref.\ \cite{L94,LS94} with regards to the stability analysis
as the string approaches the black hole singularity.
The stationary string ansatz have been studied in Ref.\ \cite{LS95}.

Finally, one can consider exact or asymptotic particular string solutions
that deserve study in its own. Asymptotic solutions in a
Friedmann--Robertson--Walker background representing highly
unstable strings have been found
\cite{GSV91}. Exact string and multistring
solutions in de Sitter spacetime have 
been found by solitonic methods (a single world-sheet describing
multiple different 
and independent strings)\cite{VMS93,CVMS94,VLS94,LS94}.
Ring solutions in black holes \cite{VE94} have also been studied.
For the sake of 
completeness it is also worth to remark that
the propagation of strings in shock wave and conical spacetimes
has been exactly solved\cite{VS90,VS'90}.

Two physical interesting quantities can be computed:
The string energy--momentum tensor integrated over
a spatial volume completely enclosing the string\cite{VS92},
at fixed time $X^0$
\begin{equation}
I^{AB}(X^0)=\int{\sqrt{-G}T^{AB}(X)d^{D-1}\vec X}~, \label{3.9}
\end{equation}
where
\begin{equation}
\sqrt{-G}T^{AB}(X)={1\over 2\pi\alpha'}\int{d\sigma d\tau\left[\dot X^A
\dot X^B-X'^AX'^B\right]\delta^{(D)}\left[X-X(\sigma, \tau)\right]}~,
\label{3.10}
\end{equation}
and the invariant string size $l$\cite{dVS93}
\begin{equation}
dl^2=-\dot X^A\dot X^BG_{AB}(X)d\sigma^2~.\label{size}
\end{equation}
[Note that the differential string size has the form of an effective
mass $m_{\text{eff}}(\tau,\sigma)$ for the geodesic motion, and actually
represents a projection from the target space onto the world sheet].

\section{Null String dynamics in Cosmological backgrounds}

Let us consider as a simple application of the formalism above the isotropic
cosmological geometry given by the conformally flat
Friedmann--Robertson--Walker metric

\begin{equation}
ds^2=R^2(X^0)\left[-(dX^0)^2+(dX^1)^2+(dX^2)^2+\cdots+(dX^{D-1})^2\right]~,
\label{c.1}
\end{equation}
where we parametrize the conformal factor as $R(X^0)=K(X^0/\beta)^\beta$.

We are interested in this paper in the null string dynamics and the
fluctuations around it.
\noindent
{\it Zeroth order:}
The equations of motion (\ref{2.6})
have the following first
integral of motion (conservation of the four-moment $P_A$)
\begin{equation}
R^2(A^0)\dot A^0(\sigma,\tau)=-P_0(\sigma)~,~~
R^2(A^0)\dot A^i(\sigma,\tau)=P_i(\sigma)~.\label{c.2}
\end{equation}
where from now on $i,j$ run from $1$ to $D-1$.

The constraints at $\tau=0$ yield
\begin{equation}
(P_0)^2=\sum_{i=1}^{D-1}(P_i)^2~,\label{c.3}
\end{equation}
\begin{equation}
-P^0(\bar{A^0})'=\sum_{i=1}^{D-1}P^i(\bar{A^i})'~.\label{c.4}
\end{equation}
where we use the notation $A^A(\sigma,\tau=0)=\bar{A^A}$

{}From Eq. (\ref{c.2}) we see that the only relevant
$\tau-$dependence is that of $A^0(\tau)$. In fact, we can write
\begin{equation}
A^i(\tau)-\bar{A^i}={P^i\over -P^0}\left[A^0(\tau)-\bar{A^0}\right]~.
\label{c.5}\end{equation}

By direct integration of the zeroth component of Eq. (\ref{c.2}) one obtains
\begin{equation}
A^0(\tau)-\bar{A^0}=\left[{(1+2\beta)\beta^{2\beta} P^0\over K^2}
\tau\right]^{1\over1+2\beta}\dot=q(\sigma)\tau^{1\over1+2\beta}~,
\label{c.6}\end{equation}

The particular time-dependence of the scale factor as usually referred to
in the literature is recalled in Table \ref{t1}. The cases dealt with in
Refs. \cite{VN92} are in $D=4$ and $\beta=-1/2$ and $-1$ respectively.
Note that the cosmic time $t$ is related to the conformal time
$X^0\approx A^0$ by
\begin{equation}
t-t_0={K\left(A^0-\bar{A^0}\right)^{1+\beta}\over\beta^\beta(1+\beta)}~.
\label{c.t}\end{equation}
Thus, $\tau$ does {\it not} coincides with $t$.

We can now compute the zeroth order energy--momentum density tensor
from Eq. (\ref{3.9})
\begin{equation}
I_A^{B}(\sigma)=\frac{P_AP_C\eta^{CB}}{-P_0\alpha'}~.
\end{equation}
Hence, at zeroth order, the string energy and momentum keep constant
as the universe evolves. The trace of $I_A^B$ vanishes
to this zeroth order. Since, to this order, each point of the string follows
a null geodesic, the proper size Eq. (\ref{size}) also vanishes.

\noindent
{\it First order fluctuations:} The $0$ component of Eq.\ (\ref{2.9})
yields
\begin{eqnarray}
&\ddot B^0+{2dR\over RdX^0}\left[\dot A^0\dot B^0+\dot A^i\dot B^i
\right]+\frac{d}{dX^0}\left({dR\over RdX^0}\right)\left[
\left(\dot A^0\right)^2+\left(\dot A^i\right)^2\right]B^0=\nonumber\cr\cr
&{A^0}''+{dR\over RdX^0}\left[
\left({A^0}'\right)^2+\left({A^i}'\right)^2\right]~.
\label{c01c}\end{eqnarray}

With the help of the constraint equation (\ref{2.10})
\begin{equation}
\left(2\dot B^A{A^B}'+\dot A^A{B^B}'\right)\eta_{AB}=0~,
\label{ca1c}\end{equation}
we obtain an uncoupled equation for $B^0$
\begin{equation}
\ddot B^0+\frac{4\alpha}{\tau}\dot B^0-\frac{2\alpha^2}{\beta\tau^2}B^0=
{A^0}''+\frac{2\left({A^i}'\right)^2}{A^0}\dot=F^0(\sigma,\tau)~,
\label{c01cdes}\end{equation}
where $\alpha=\beta/(\beta+1)$, and we have used that
\begin{equation}
\frac{dR}{RdX^0}=\frac{\beta}{X^0}~,~~
\frac{d}{dX^0}\left(\frac{dR}{RdX^0}\right)=\frac{-\beta}{\left(X^0\right)^2}~.
\label{derfaces}\end{equation}

We can write the general solution to Eq. (\ref{c01cdes}) as
\begin{equation}
B^0(\sigma,\tau)=\sum_k b_k(\sigma)\tau^k
\label{serie}\end{equation}
where $b_{k_+}$ and $b_{k_-}$ are arbitrary constants corresponding to the
general solution to the homogeneous equation associated to Eq. (\ref{c01cdes});
$k_\pm=-(4\alpha-1)/2\pm\sqrt{(2\alpha-1/2)^2+2\alpha^2/\beta}$. The other
powers of $k$ correspond to the series development of the function
$F^0(\sigma,\tau)$ and give the particular solution to the inhomogeneous
equation.

To provide a simpler solvable example we take
$\bar{A^0}=0$. In this case $F^0$ takes the following form:
\begin{equation}
F^0(\sigma,\tau)=\left[q''+\frac{2\beta}{q}\left({P_iq\over P_0}
\right)'^{~2}\right]\tau^{\alpha/\beta}
-\frac{4\beta{\bar{{A}^i}}'}{q}\left({P_iq\over P_0}\right)'
+\frac{\beta\left({\bar{{A}^i}}'\right)^2}{q}\tau^{-\alpha/\beta}~.
\label{f0}\end{equation}

Plugging this into the right hand side of Eq. (\ref{c01cdes}) and using the
form of $B^0$ given by Eq. (\ref{serie}) we obtain
\begin{equation}
\sum_kb_k[k(k-1)+4k\alpha-2\alpha^2/\beta]\tau^{k-2}=F^0(\sigma,\tau)
\label{eqf0}\end{equation}

Thus, there are only three values of $k$ given by $k-2=0,\pm\alpha/\beta$
with the corresponding $b_k$ as defined by Eqs.\ (\ref{f0}) and
(\ref{eqf0})
\begin{eqnarray}
b_2&=&-\frac{4\beta{\bar{{A}^i}}'}{q[2+8\alpha-2\alpha^2/\beta]}
\nonumber\\
b_{2+\alpha/\beta}&=&{q''+\frac{2\beta}{q}\left({P_iq\over P_0}
\right)'^{~2}\over(2+\alpha/\beta)(1+4\alpha+\alpha/\beta)
-2\alpha^2/\beta}\label{bks}\\
b_{2-\alpha/\beta}&=&\frac{\beta\left({\bar{{A}^i}}'\right)^2}
{q\left[(2-\alpha/\beta)(1+4\alpha-\alpha/\beta)-2\alpha^2/\beta\right]}
\nonumber
\end{eqnarray}

Let us now consider the $i$ components of Eq. (\ref{2.9}) for the first
order fluctuations:

\begin{equation}
\ddot B^i+\frac{2}{R}\frac{dR}{dX^0}\left[\dot A^0\dot B^i+\dot A^i\dot B^0
\right]+2\frac{d}{dX^0}\left(\frac{dR}{RdX^0}\right)\dot A^0\dot A^iB^0=
{A^i}''+2\frac{dR}{RdX^0}{A^0}'{A^i}'~.
\label{ci1c}\end{equation}

With the help of Eq. (\ref{derfaces})
\begin{equation}
\ddot B^i+\frac{2\alpha}{\tau}\dot B^i=
{A^i}''+\frac{2\beta{A^i}'{A^0}'}{A^0}+\frac{2\alpha P_i}{P_0\tau}
\left(\dot B^0-\frac{\alpha B^0}{\beta\tau}\right)
\dot=F^i(\sigma,\tau)~,
\label{ci1csimp}\end{equation}

Known $B^0(\sigma,\tau)$ the general solution to this equation can be
written in terms of quadratures
\begin{equation}
B^i(\sigma,\tau)=\bar{B^i}+\bar{\dot{B^i}}
\frac{\tau^{1-2\alpha}}{1-2\alpha}+
\int^\tau d\tilde{\tilde\tau}\tilde{\tilde\tau}^{-2\alpha}
\int^{\tilde{\tilde\tau}}
 d\tilde\tau{\tilde\tau}^{2\alpha}F^i(\sigma,\tilde\tau)~,
\label{cuadraturas}\end{equation}
where $\bar{B^i}=B^i(\sigma,0)$. Note that $\bar{B^i}$ and $\dot{\bar{B^i}}$
can be absorbed into $\bar{A^i}$ and $P_i$ respectively. We are thus left
with the two constants $b_{k_\pm}$ which are fixed by the two first order
constraints equations (\ref{2.10}) and (\ref{2.11})
\begin{eqnarray}
&\left(\dot A^A\dot B^B+{A^A}'{A^B}'\right)\eta_{AB}=0~,\cr
&\left(2\dot B^A{A^B}'+\dot A^A{B^B}'\right)\eta_{AB}=0~.
\label{vinsim}\end{eqnarray}
We are thus sure we are not adding any spurious degree of freedom at first
order.

We can again show the explicit dependence for the case $\bar{A^0}=0$. 
In this case $F^i$ takes the following form
\begin{eqnarray}
&F^i(\sigma,\tau)=\bar{A^i}''+2b_2\alpha{P_i\over P_0}(2-\alpha/\beta)+
2\beta{q'\over q}\bar{A^i}'-\cr\cr
&\left[\left({P_iq\over P_0}\right)''+
2\beta{q'\over q}\left({P_iq\over P_0}\right)'-4\alpha b_{2+\alpha/\beta}
{P_i\over P_0}\right]\tau^{\alpha/\beta}+\cr\cr
&4\alpha b_{2-\alpha/\beta}{P_i\over P_0}(1-\alpha/\beta)
\tau^{-\alpha/\beta}+2\alpha b_{k_\pm}{P_i\over P_0}
\left((k_\pm)-\alpha/\beta\right)\tau^{k_\pm-2}~,
\label{fi}\end{eqnarray}
or in a more compact way
\begin{equation}
F^i(\sigma,\tau)=\sum_n F^i_n(\sigma)\tau^n~,~~n=0,
\pm\alpha/\beta,(k_\pm)-2~,
\end{equation}
where the notation $F^i_n$ clearly refers to the coefficient of
the $n$-th power of $\tau$ in Eq. (\ref{fi}).

Plugging this into the right hand side of Eq. (\ref{cuadraturas})
and using the form of $B^0$ given by Eqs.\ (\ref{serie}) and
\ (\ref{bks}), we obtain
\begin{equation}
B^i(\sigma,\tau)=\bar{B^i}+\bar{\dot{B^i}}
\frac{\tau^{1-2\alpha}}{1-2\alpha}+
\sum_n {F^i_n\tau^{n+2}\over(n+2)(2\alpha+n+1)}~~,~~~
n=0,\pm\alpha/\beta,(k_\pm)-2~.
\label{eqbi}\end{equation}

\bigskip
The proper string size that vanished to zeroth
order now takes the following form
\begin{equation}
dl^2=R^2\eta_{AB}{A^A}'{A^B}'d\sigma^2~,
\label{size1}\end{equation}
where we have used the constraint (\ref{vinsim}). We see that the string
size grows like the conformal factor of the
Friedmann--Robertson--Walker metric. This unstable
behavior was already observed in Ref. \cite{SV90} in the 
fluctuations around the center of mass zeroth order solution.

The coordinate fluctuations $B^0$ and $B^i$ evolve with several powers
of $\tau$ and for a given value of the conformal factor exponent $\alpha$,
we will have several regimes. We can see from the above expressions
(Eq.\ (\ref{eqbi})) that
in the regime of $R$ large, that is
$\tau\to\infty$, for $\alpha/\beta>0$, i.e. $\beta>-1$, the
overwhelming power will be $(2+\alpha/\beta)$ or $(2+k_+)$. When
$\alpha/\beta<0$, i.e. $\beta<-1$, the regime of $R$ large is reached
when $\tau\to0$,
then the predominant power will be $2+\alpha/\beta$ or $2+k_-$. Note also
that these results do not depend explicitly on the dimension, $D$, since
the metric coefficients neither do.

\section{Black Hole spacetimes}

We are particularly interested in studying the dynamics of strings in black
hole spacetimes and its approach to the singularity at radial coordinate $r=0$.

Let us consider a $D$--dimensional black hole background \cite{MP86},
\begin{equation}
ds^2=-g(r)(dX^0)^2+g(r)^{-1}dr^2+r^2d\Omega_{D-2}~, \label{3.1}
\end{equation}
where for the Reissner-Nordstr\"om metric 
\begin{equation}
g(r)=1-\left({R_S\over r}\right)^{D-3}+\left({\tilde Q\over r}
\right)^{2(D-3)}~,~~\tilde Q^{2(D-3)}={8\pi GQ^2\over(D-2)(D-3)}~,
\end{equation}
with $M$ and $Q$ being respectively the mass and charge of the black hole.

We can now study the string dynamics by using the null string approach
described in Sec.\ II.

We then propose the following development for the string
coordinates
\begin{equation}
X^A(\sigma, \tau)=A^A(\sigma, \tau)+c^2B^A(\sigma, \tau)+...
\end{equation}
Replacing this into the field equations (\ref{2.2}), with (\ref{3.1})
acting as the 
curved background, we find to zeroth order the following first integrals
of motion
\begin{equation}
\dot\theta\simeq \dot A^2=0~,~~\dot\phi\simeq
A^3={L(\sigma)\over r^2}~,\label{3.2'}
\end{equation}
where for simplicity we consider a string lying on the equatorial
plane. $L(\sigma)$ is the angular momentum of each point forming the string.
The other first integrals of motion are
\begin{equation}
\dot X^0\simeq \dot A^0=E(\sigma)g^{-1}(r)~,~~
\dot r^2\simeq (\dot A^1)^2=
E^2(\sigma)-{L^2(\sigma)\over r^2}g(r)~.\label{3.2}
\end{equation}
From this last equation one can see that for the Reissner-Nordstr\"om 
black hole, if $L(\sigma)\not=0$, the string
will only skirt the singularity. Since we are interested in the
string approach to the singularity, we
consider from now on Schwarzschild black holes (in $D$ dimensions).

The remaining constraint equation
(\ref{2.8}) at $\tau=0$ takes the following simple form
\begin{equation}
\bar{A^0}'(\sigma)={L(\sigma)\over E(\sigma)}\bar{A^3}'(\sigma)~.
\end{equation}

We can further integrate expression (\ref{3.2}) near the black
hole singularity to obtain
\begin{equation}
A^1(\sigma,\tau)\simeq p(\sigma)(-\tau)^{2\over D+1}~,~~
p(\sigma)=[{1\over2}(D+1)R_S^{D-3\over 2} L(\sigma)]^{2\over D+1}
\label{3.3}
\end{equation}
where we have chosen $\tau\leq0$ such that $A^1(\tau=0)=0$.
$\tau$ measures the proper time of infall to the singularity at $r=0$.

By use of the above expression we find
\begin{equation}
A^0(\sigma,\tau)\simeq \bar{A^0}+{(D+1)\over(3D-5)}E
\left({p(\sigma)\over R_S}\right)^{D-3}(-\tau)^{3D-5\over D+1}~,
\label{3.4}
\end{equation}
and
\begin{equation}
A^3(\sigma,\tau)\simeq\bar{A^3}-{(D+1)\over (D-3)}
{L(\sigma)\over p^2(\sigma)}(-\tau)^{D-3\over D+1}~,\label{3.5}  
\end{equation}

\begin{equation}
A^i=\pi/2 {\text{ for}}~~ i=2,4,5,..., D-1
\end{equation}

Comparison of these expressions with the
corresponding ones found in Ref \cite{LS93},  (Eq. (88)),
shows that they
have the same $\tau$ dependence. The difference lies in the
prefactors: In Eqs. (\ref{3.3})--(\ref{3.4}) they are
$\sigma$--dependent while in those of Ref \cite{LS93} they
are constant. 
This is due to the different zeroth order solution
we have chosen. In Ref.\ \cite{LS93}
we have taken the center of mass motion which is a point--like
solution, while now we have a null string, an extended zeroth order
solution.

We can consider the approach to the singularity with
higher powers of $A^1$ (our radial coordinate). This yields
the following power dependence in $\tau$
\begin{equation}
A^1=\sum_{n=1}^\infty A^1_{n}(\sigma)(-\tau)^{2n/(D+1)}~,
\end{equation}
and then
\begin{equation}
A^3=\sum_{n=1}^\infty A^3_{n}(\sigma)(-\tau)^{(D-5+2n)/(D+1)}~,
\end{equation}
\begin{equation}
A^0=\sum_{n=1}^\infty A^0_{n}(\sigma)(-\tau)^{1+2n(D-3)/(D+1)}~.
\end{equation}

This $\tau$ dependence seems to be a quite general behavior near
singularities and generalize to $D$--dimensions the case analyzed
recently in Ref (\cite{VE95}). There will still be higher order
corrections coming form the first
order fluctuations $B^\mu$, second order fluctuations $C^\mu$,
and so on. We have already seen that in the cosmological case,
five different powers of $\tau$ contribute to the first order.

The proper string size that at zeroth order vanished for
null string, does not vanish at higher order.
In order to compute the first order correction to the string
size we do not need to explicitly compute
the $B^A$ terms. In fact, by use of the constraint (\ref{2.4}),
Eq. (\ref{size}) can be rewritten as
\begin{equation}
\left({dl\over d\sigma}\right)^2={X^A}' {X^B}'G_{AB}(X)~.\label{size'}
\end{equation}
Plugging expressions (\ref{3.2})--(\ref{3.2'}) into this equation and
making use of the constraint (\ref{2.8}) to eliminate ${A^1}'$, we obtain
\begin{equation}
\left({dl\over d\sigma}\right)^2=-g({A^0}')^2+ (A^1)^2({A^3}')^2+ 
g{(E{A^0}'-L{A^3}')^2\over(E^2-L^2g/(A^1)^2)}~.\label{size''}
\end{equation}
As the string approachs the singularity it diverges in the following way
\begin{equation}
\left({dl\over d\sigma}\right)^2\to\left({R_S\over r}\right)^{D-3}
(\bar{X^0}')^2\to(-\tau)^{-2(D-3)/(D+1)}~.\label{size'''}
\end{equation}

This unstable behavior have already been discovered in Ref \cite{LS93},
where we studied string propagation in $D$--dimensional Reissner--Nordstr\"om 
black holes under the approximation labeled by $c=1$.
We solved there the first order
string fluctuations around the center of mass motion at spatial infinity, 
near the event horizon and at the spacetime singularity.
In Reissner--Nordstr\"om black holes the
radial components and angular string components
develop instabilities as the string approachs the singularity.
The string motion is like that of a particle in a strongly
attractive potential proportional to 
$-(\Delta\tau)^{-2}$. As $\Delta\tau\to 0$ the string ends trapped by the
black hole singularity.

We can also compute the integrated energy--momentum tensor from expressions
(\ref{3.9}) and (\ref{3.10}). Using Eqs. (\ref{3.2})--(\ref{3.2'}) we find
\begin{equation}
\Theta^{AB}={dI^{AB}\over d\sigma}\approx {1\over2\pi\alpha'}
{\sqrt{-g}\dot A^A\dot A^B\over\sqrt{E^2-L^2g/(A^1)^2}}~,
\end{equation}
where we have taken into account that inside the black hole $A^0$ is
spacelike and $A^1$ is timelike.

In the limit $r\to0$, we have
\begin{equation}
\Theta^1_1\to-{L\over2\pi\alpha'r}=-\Theta^3_3~,~~~\Theta^0_0\to0~.
\end{equation}
We first observe that the results are independent of the spacetime
dimensionality, $D$. The facts that the trace vanishes
(since we are dealing with a null system),
and that $\Theta^0_0\to0$, reduce the system to a $D-1$--dimensional
null system (in our example of a planar string, this reduces
the system to a two dimensional null string).

\section{Exact string solutions inside a black hole}

In four dimensions the equations of motion of a string in the curved
background (\ref{3.1}) read \cite{LS95}

\begin{eqnarray}
\ddot{t}&-&t''+\frac{g_{,r}}{g}(\dot{t}\dot{r}-
t'r')=0,\nonumber\\
\ddot{r}&-&r''-\frac{g_{,r}}{2g}(\dot{r}^2-r'^2)+
\frac{gg_{,r}}{2}(\dot{t}^2-t'^2)-gr(\dot{\theta}^2-\theta'^2)-
gr\sin^2\theta
(\dot{\phi^2}-\phi'^2)=0,\nonumber\\
\ddot{\theta}&-&\theta''+\frac{2}{r}(\dot{\theta}
\dot{r}-\theta' r')-\sin\theta\cos\theta(\dot{\phi}^2-\phi'^2)=0,\nonumber\\
\ddot{\phi}&-&\phi''+\frac{2}{r}(\dot{\phi}
\dot{r}-\phi'r')+2\cot\theta(\dot{\theta}\dot{\phi}-\theta'\phi')=0,
\label{ecmov}\end{eqnarray}
and the constraints

\begin{eqnarray}
&-&g\dot{t}t'+
\frac{1}{g}\dot{r}r'+r^2\dot{\theta}
\theta'+r^2\sin^2\theta\dot{\phi}\phi'=0,\nonumber\\
&-&g(\dot{t}^2+t'^2)+\frac{1}{g}(\dot{r}^2+r'^2)
+r^2(\dot{\theta}^2+\theta'^2)+r^2\sin^2\theta(\dot{\phi}^2+\phi'^2)=0.
\label{vinculos}\end{eqnarray}

We are interested in finding solutions inside the black hole, where we
know that the $r$ coordinate becomes timelike while the $t$ coordinate
is spacelike. We thus look for solutions of the following particular
form
\begin{equation}
r=r(\tau)\dot={\cal T}(\tau) \quad{\text{and}}\quad
t=t(\sigma)\dot={\cal R}(\sigma)~.
\label{ansatz}\end{equation}

Using this ansatz the first equation of motion in (\ref{ecmov}) give us
\begin{equation}
{\cal R}(\sigma)=E\sigma+t_0~,
\label{tdesigma}\end{equation}
where $E$ and $t_0$ are arbitrary constants.

From the first constraint Eq.\ (\ref{vinculos}) we have
\begin{equation}
{\cal T}(\tau)^2\sin^2\theta\dot{\phi}\phi'=0~.
\label{vincuno}\end{equation}

In terms of Kruskal coordinates we have
\begin{eqnarray}
u(\sigma,\tau)&=&\left(1-{{\cal T}(\tau)\over2M}\right)^{1/2}
e^{{\cal T}(\tau)\over4M}\sinh\left({{\cal R}(\sigma)\over4M}\right)\\
v(\sigma,\tau)&=&\left(1-{{\cal T}(\tau)\over2M}\right)^{1/2}
e^{{\cal T}(\tau)\over4M}\cosh\left({{\cal R}(\sigma)\over4M}\right)
\end{eqnarray}

Eq.\ (\ref{vincuno}) allows us to analyze three cases:

\par\noindent
a) $\theta=\text{constant}\not=\pi/2$. Then $\phi$ is also constant and
\begin{equation}
\dot {\cal T}(\tau)=\pm Eg({\cal T})
\label{ra}\end{equation}
 represents a straight string in the
two dimensional ${\cal R}$--$\phi$ space.
\par\bigskip\noindent
b) For $\theta=\pi/2$, i.e. the planar case, we have:

\par\noindent
b1) $\dot{\phi}=0$. Then, from the equation of motion
of $\phi$ we deduce $\phi=n\sigma+\phi_0$,
where $n$ and $\phi_0$ are arbitrary constants.
While for the timelike coordinate ${\cal T}(\tau)$ we have
\begin{equation}
\dot {\cal T}(\tau)^2=E^2g({\cal T})^2-n^2{\cal T}(\tau)^2~.
\label{rb}\end{equation}
This solution represents a straight twisted string in
the ${\cal R}$--$\phi$ space.

\par\noindent
b2) $\phi'=0$. Then, from the equation of motion
of $\phi$ we deduce $\dot\phi=L/{\cal T}(\tau)^2$. Here $L$
is a constant and now
\begin{equation}
\dot {\cal T}(\tau)^2=E^2g({\cal T})^2-
g({\cal T}){L^2\over {\cal T}(\tau)^2}~.
\label{rc}\end{equation}
This solution represents a rigid rotating 
straight string in the ${\cal R}$--$\phi$ space.

\section{Conclusion}

We have performed a systematic study of the null string expansion
method to solve the string equations of motion and constraints in
curved spacetimes. This corresponds to a small string tension
expansion (inverse powers of $\alpha'$). The perturbative 
expansion series is conveniently described in terms of the 
parameter $c$, the wave propagation velocity along the string.
The zeroth order describes a null string configuration; first and
higher order fluctuations around it are systematically constructed.
The null string expansion series in $D$--dimensional 
Friedmann--Robertson--Walker and Schwarzschild spacetimes have been
computed. In the latter case, the first terms of the series
represent a generic approach to the string falling towards the
$r=0$ singularity. The string integrated energy-momentum tensor
and string proper size have been computed and analyzed.

The different methods to solve the string dynamics in curved
apacetime have been characterized in terms of the parameter $c$
(or equivalently, in terms of the ratio between $\dot X^A$ and
$c{X^A}'$), as summarized in Sec.\ II and Table\ II:

(i) ``Impulsive'', $c\ll1$ (null string expansion);

(ii) ``Perturbative'', $c=1$ (center of mass expansion);

(iii) ``Adiabatic'', $c\gg1$ (dual to (i), initial configuration
expansion);

(iv) ``Composite'', ($\tau\to0$ expansion, ring string expansion).

The null string expansion is well appropriate to describe the
string propagation in the strong gravitational
field regime and well encompasses
the $\tau\to0$ expansion in cosmologiacal and black hole spacetimes,
in the string unstable regime where the proper size of the
string grows like the scalar factor, and like 
$({R_S/r})^{(D-3)/2}\sim(-\tau)^{-(D-3)/(D+1)}$
inside the Schwarzschild black hole (near the $r=0$ singularity).

\begin{acknowledgments}
C.O.L was supported by the NSF grant PHY-95-07719 and by research
founds of the University of Utah.
He also thanks the DEMIRM--Observatoire de Paris, where part of this
work was done, for kind hospitality and use of its working facilities.
\end{acknowledgments}

\begin{table}
\caption{Here we recall the explicit conformal time ($\eta$)
and cosmic time ($t$) --dependence
of the flat Friedmann--Robertson--Walker
metric as usually referred to in the literature
\protect\cite{GSV91}.}
\bigskip
\begin{tabular}{ccl}
$R(\eta)=(\eta/\beta)^\beta$&$R(t)\sim t^{\alpha=\beta/(\beta+1)}$
&Spacetime\\
\hline
constant, $\beta\to0$&constant&Flat space\\ 
$\eta^\beta$, $0<\beta<\infty$&$t^{\alpha}$, $0<\alpha<1$&
Standard cosmology\\
$e^\eta$, $\beta\to\infty$&$t$&Linear expansion\\
$\eta^\beta$, $-1<\beta<0$&$t^{\alpha}$, $-\infty<\alpha<0$&
Super inflation\\
$\eta^{-1}$&$e^{Ht}$&de Sitter\\
$\eta^\beta$, $-\infty<\beta<-1$&$t^{\alpha}$,
$1<\alpha<\infty$&Power-law inflation\\
\end{tabular}
\label{t1}
\end{table}

\begin{table}
\caption{In this table we show the different approaches to
study the string dynamics in curved
spacetimes. $c\ll 1$, $c=1$ and $c\gg 1$ correspond to 
$\partial_{\tau}X^A\gg c\partial_{\sigma}X^A$, $\partial_{\tau}X^A\sim
c\partial_{\sigma}X^A$ and $\partial_{\tau}X^A\ll c\partial_{\sigma}X^A$
respectively. Hybrid cases are still possible when some components of
the string coordinate $X^A$ satisfy eventually one of the inequalities
and other coordinate components do not. The regimes of stability of
the first order fluctuations are also shown as well as the references
in which the approaches (in Cosmology and Black Holes) have been
originally developped.}
\bigskip
\begin{tabular}{lcccc}
&\multicolumn{2}{c}{Black Holes}&\multicolumn{2}{c}{Cosmology}\\
\tableline
Approach&Zeroth Order&Fluctuations&Zeroth Order&
Fluctuations\\
\ &\ &$R\to0$&\ &$R\to\infty$\\
\tableline
(i) Impulsive, $c\ll1$&null string\tablenotemark[1]
&unstable&null string\tablenotemark[2]&unstable\\
\ \ \ \ \ \ $\dot X^A\gg cX^A{'}$&\ &\ &\ &\ \\
(ii) Perturbative, $c=1$&center of mass\tablenotemark[3]
&unstable&center of mass\tablenotemark[4]&stable/unstable\\
\ \ \ \ \ \ $\dot X^A\sim cX^A{'}$&\ &\ &\ &\ \\
(iii) Adiabatic, $c\gg1$&initial configuration\tablenotemark[1]
&unstable&initial configuration\tablenotemark[1]&unstable\\
\ \ \ \ \ \ $\dot X^A\ll cX^A{'}$&\ &\ &\ &\ \\
(iv) Hybrid&ring\tablenotemark[5]&unstable&
asymptotic\tablenotemark[6]&stable/unstable\\
\end{tabular}
\tablenotetext[1]{This paper}
\tablenotetext[2]{Ref.\ \cite{VN92}.}
\tablenotetext[3]{Ref.\ \cite{LS93}.}
\tablenotetext[4]{Ref.\ \cite{VS87}.}
\tablenotetext[5]{Ref.\ \cite{L94}.}
\tablenotetext[6]{Ref.\ \cite{GSV91}.}
\label{t2}
\end{table}
\end{document}